\documentclass[runningheads]{llncs}

\usepackage{booktabs}   
\usepackage{subcaption} 
\usepackage{cleveref}

\usepackage{listings}
\usepackage{lstlangcoq}

\usepackage{bussproofs}

\usepackage[export]{adjustbox}
\usepackage[normalem]{ulem}
\usepackage{semantic}
\usepackage{mathpartir}
\usepackage{txfonts}
\usepackage{iris-llncs}
\usepackage{url}

\lstdefinestyle{customc}{
  belowcaptionskip=1\baselineskip,
  breaklines=true,
  xleftmargin=\parindent,
  language=C,
  showstringspaces=false,
  basicstyle=\ttfamily,
  keywordstyle=\bfseries\color{green!40!black},
  commentstyle=\itshape\color{purple!40!black},
  identifierstyle=\color{blue!50!black},
  stringstyle=\color{orange},
  numbers=left,
}

\lstdefinestyle{customcoq}{
  mathescape=true,
  belowcaptionskip=1\baselineskip,
  breaklines=true,
  xleftmargin=\parindent,
  language=Coq,
  morekeywords={Variant, fun},
  emph={%
    SOCKAPI,ITree,data_at,data_at_
  },
  emphstyle={\bfseries\color{green!40!red!80}},
  showstringspaces=false,
  basicstyle=\small\ttfamily,
  keywordstyle=\bfseries\color{green!40!black},
  commentstyle=\itshape\color{purple!40!black},
  identifierstyle=\color{violet!80!black},
  stringstyle=\color{orange},
   literate={=>}{{$\Rightarrow$\ }}1
   {-}{{\textsf{-}}}1
   {->}{{$\rightarrow\,$}}1
   {-->}{{$\longrightarrow\,$}}2
   {<->}{$\leftrightarrow$}1
   {<-->}{$\!\longleftrightarrow\,$}1
   {>=}{$\!\ge\,$}1
   {>=>}{$\subtype$}1
   {<=>}{$\Leftrightarrow\,$}2
   {<=}{$\!\le\,$}1
   {forall}{$\forall\,$}1
   {forallb}{forallb}7
   {existsb}{existsb}7
   {existsv}{existsv}7
   {\/\\}{{$\wedge$\ }}1
   {|-}{{$\,\vdash\,$}}2
}

\newcommand{\inlinecoq}[1]{\lstinline[style=customcoq,columns=flexible]{#1}}

\newcommand{\ignore}[1]{}

\newcommand{\pk}{\mathit{pk}}
\newcommand{\pv}{\mathit{pv}}
\newcommand{\islock}{\boxdotright}

\begin{document}
\title{Bringing Iris into the Verified Software Toolchain} 

\author{William Mansky\inst{1}}

\authorrunning{W. Mansky} 

\institute{University of Illinois Chicago, USA}

\maketitle

\begin{abstract}
The Verified Software Toolchain (VST) is a system for proving correctness of C programs using separation logic. By connecting to the verified compiler CompCert, it produces the strongest possible guarantees of correctness for real C code that we can compile and run. VST included concurrency from its inception, in the form of reasoning about lock invariants, but concurrent separation logic (CSL) has advanced by leaps and bounds since then. In this paper, we describe efforts to integrate advancements from Iris, a state-of-the-art mechanized CSL, into VST. Some features of Iris (ghost state and invariants) are re-implemented in VST from the ground up; others (Iris Proof Mode) are imported from the Iris development; still others (proof rules for atomic operations) are axiomatized, with the hope that they will be made foundational in future versions. The result is a system that can prove correctness of sophisticated concurrent programs implemented in C, with fine-grained locking and non-blocking atomic operations, that yields varying soundness guarantees depending on the features used.
\end{abstract}


\section{Introduction}
Soon after O'Hearn observed that separation logic could be naturally extended to concurrency \cite{concurrent-sep}, Gotsman et al.~\cite{gotsman} and Hobor et al.~\cite{oracle-semantics} independently came up with variants that used C-style dynamically created locks and threads, where each lock was associated with a \emph{lock invariant} describing the resources protected by the lock. Hobor et al.'s variant was incorporated into the Verified Software Toolchain (VST)~\cite{vst}, a system for proving correctness of C programs in Coq using separation logic, giving it the power to verify (well-synchronized) concurrent C programs. This power then went largely unused for several years. In the meantime, there was a boom in sophisticated concurrent separation logics, including FCSL~\cite{FCSL-hist}, TaDA~\cite{tada}, and Iris~\cite{iris}, that could prove more interesting correctness properties and address more interesting synchronization patterns (especially fine-grained and/or lock-free patterns). Iris in particular synthesized several streams of prior work with the claim that ``ghost states and invariants are all you need'', allowing it to define features like the atomic Hoare triples of TaDA in terms of more primitive constructs (even invariants turned out to be expressible using ghost state~\cite{iris-ground-up}). Iris has since been the focus of many exciting developments in CSL, including a custom proof mode that provides separation-logic analogues to Coq tactics~\cite{mosel}, a derived logic for reasoning about relaxed-memory atomics~\cite{iris-weak}, a proof system for the core language of Rust~\cite{rustbelt}, and even a refinement type system for verifying C programs~\cite{refinedc}.

The fundamental logic of Iris is language-agnostic, but it is usually applied to simple functional programming languages extended with features of interest (weak memory models, lifetimes, communication primitives, etc.). RefinedC~\cite{refinedc}, which includes an extensive C semantics, is a notable exception, but its semantics is not yet validated by any other C-related tools, and it forgoes tactics for reasoning about C code in favor of fully automated type checking. If we want the strong guarantees of VST's connection to the CompCert verified compiler~\cite{compcert}, or need to do interactive verification beyond the reach of automated type systems, we may still want to take advantage of VST's infrastructure. This led us to ask: how much of the separation logic innovation of Iris can be used in VST? Where do the foundations overlap, and where do they differ? Where can results be transferred directly, and where do they need to be adapted? The rest of this paper describes the answers. In summary:

\begin{enumerate}
\item The foundational logical model of Iris is different from that of VST. In particular, in Iris everything down through the basic points-to assertion is ghost state, describing logical ownership of portions of a monolithic physical state. VST takes a more traditional approach in which the points-to assertion is primitive, and the notion of ownership of memory by threads persists through the CompCert proofs down to assembly. Thus, to begin, \textbf{we extend the heaps of VST with Iris-style ghost state.} Once we have done this, we can recapitulate the construction of ghost state assertions and global invariants in VST.

\item Most of the client-facing features of Iris operate not on the logic of Iris per se, but on a typeclass of separation logics that describes the basic rules most such logics provide. The logic of Iris is then proved to be an instance of this typeclass. \textbf{We prove that VST's logic is another instance of this typeclass, and then immediately gain the ability to use Iris Proof Mode in VST proofs, without changing the behavior of any of VST's tactics.} This is a remarkable illustration of the flexibility of the Iris development, although it comes with some caveats: VST supports only a significantly less useful version of the ``persistent'' modality, and its symbolic execution engine is not compatible with that of Iris. Even so, Iris Proof Mode is very useful for proving complex separation logic implications, such as commonly arise when reasoning with invariants.

\item Unlike the lock invariants of VST, which we only interact with when acquiring or releasing locks, the global invariants of Iris can be accessed at any time between steps of a program, and also during the execution of atomic operations---either physically atomic operations (reads, writes, etc.) or more complex functions proved to be \emph{logically atomic}. In C, ordinary reads and writes cannot be assumed to be atomic: data races on them have undefined behavior, allowing compilers and hardware to aggressively optimize them. For atomic access to shared memory, C11 includes a syntax for designated atomic memory operations. \textbf{We axiomatize the behavior of the C11 atomics as instances of Iris's rule for physical atomics, and use this to prove correctness of programs that use C11 atomics in sequentially consistent mode. We can also use the same pattern to define logically atomic triples in VST, and use them to prove the most general specifications for C implementations of linearizable data structures.} The axioms added here are not yet proved sound down to assembly: the proof connecting VST's original lock operations to the correctness guarantees of CompCert is already a considerable and unfinished effort~\cite{cpm}, and would have to be further extended to handle atomic operations. Still, the work of Lahav et al.~\cite{rc11} gives us confidence that the axiomatization is correct, and we hope to complete the proof in future work.
\end{enumerate}

The work described here does not allow us to take advantage of every innovation in Iris, nor does it give us guarantees down to assembly for every sophisticated concurrent program we are able to verify, but it goes a long way towards bringing VST up to date with the latest tools in concurrent separation logic. It also lays a path for adding further Iris features to VST (for instance, we could imagine axiomatizing relaxed-memory atomics along the same lines as our SC atomics, though their soundness question is even trickier), and should provide some guidance to developers of other systems who want to take advantage of the latest developments in Iris. In the rest of this paper, we will describe each of these steps in detail, and show examples of the C programs we are now able to verify in VST. The features described are present in VST as of version 2.8, released in June 2021.

\section{Ghost State in VST}
\label{VST-ghost}

\subsection{Adding Ghost State to the Model}
An assertion in separation logic is ultimately a predicate on some model of program state. In a simple logic, the model may be just a map from memory addresses to values, but this model quickly becomes insufficient when we want to add interesting features like fractional ownership or ``predicates in the heap'' (used for function pointers and invariants). Instead, most separation logics use some form of annotated or extended heap model, and then erase the annotations when relating proofs to the behavior of actual programs. The model used by VST is based on the work of Hobor et al.~\cite{knot}, and takes the form of a map from locations to resources that include share information, lock invariants, and function pointers as well as values. In Iris, on the other hand, the actual physical memory is almost completely absent from the model: assertions are interpreted on a generic ``camera'' that includes all the kinds of ghost state used to verify a particular program, which always includes a piece set aside to track ownership of heap locations. The ``erasure'' then throws away the entire model, leaving only the monolithic physical state (the complete map of locations to concrete values).

Though re-engineering VST's model to more closely match that of Iris would be an interesting challenge, it turns out to be important to VST's soundness proof that we do not erase ownership of heap locations when we erase the annotations. In the proofs that lift CompCert's sequential compiler correctness properties to a concurrent setting~\cite{cpm}, annotations are erased in the first step (to the ``juicy'' machine), while ownership is translated into CompCert \emph{permissions} and used to guarantee race-freedom in the lower levels of the soundness proof. As a result, moving VST to an Iris-style model while retaining correctness guarantees down to assembly may be impossible, and is certainly beyond the scope of this paper. Instead, we extend VST's model to include Iris-style ghost state as an additional component, and re-derive laws for manipulating it that are almost (but not quite) identical to those of Iris. The foundational ghost state described in this section has been part of VST for several years, and is used in developments such as the connection to CertiKOS by Mansky et al.~\cite{connecting}.

The program logic of VST, called Verifiable C, is proved sound with respect to a semantic model in which 
assertions are modeled as predicates over \emph{resource maps}, of type \inlinecoq{rmap}.
An \inlinecoq{rmap} is a \emph{step-indexed} map from CompCert memory locations to the associated separation-logic values: either partial ownership of a value in memory, or a ``predicate in the heap'' representing a separation logic assertion, such as a global assertion (e.g., the specification of a function pointer) or a lock invariant. The step index breaks the circularity in the mutual dependence between assertions and \inlinecoq{rmap}s; all assertions must be preserved under a decrease in step index, and the step index serves as an ``approximation level'' for all the assertions in the \inlinecoq{rmap}~\cite{vst}.

As described by Jung et al.~\cite{iris-ground-up}, a collection of disjoint ghost states drawn from a range of different algebras\footnote{In separation logics with generic ghost state, each piece of ghost state is drawn from a particular algebra describing the ways in which that resource can be combined and manipulated. Depending on the logic, these algebras may be called ``resource algebras'', ``partial commutative monoids'', or various other names; for clarity, we refer to them as ``ghost algebras'' throughout.} can be combined into a single, top-level ghost state $M$, a finite map from each of a set of \emph{indices} to an element of the ghost algebra at that index. We add ghost state to the \inlinecoq{rmap} type precisely by adding such a map: where \inlinecoq{rmap} was originally a step-indexed construction applied to the type \inlinecoq{address->resource}, the same construction is now applied to the type \inlinecoq{(address->resource) * ghost}, where \inlinecoq{ghost} is a finite map from indices (natural numbers) to dependent pairs $(G, \{a : \mathsf{car}(G)\ |\ \mathsf{valid}\ a\})$ of a ghost algebra $G$ and a valid element $a$ of that algebra. While in Iris the map from indices to ghost algebras is itself a ghost algebra, the construction we use in VST is not sophisticated enough to prove this without universe inconsistencies; as we will see below, this limits the kinds of higher-order ghost state we can define in VST, although it is sufficient for the common case of the \emph{agreement algebra} used to derive invariants (see section~\ref{invariants}).

Next, we must show that our augmented \inlinecoq{rmap}s are still a suitable model for separation logic. Fortunately, VST's core separation logic (of which Verifiable C is an instance) is designed to be agnostic to its language and model: in particular, it is applicable to any resource that is a \emph{separation algebra}~\cite{sep-algebra}. More precisely, following the terminology of Cao et al.~\cite{sep-jungle}, the underlying separation algebra is assumed to be \emph{associative}, \emph{commutative}, \emph{functional}, \emph{positive}, and \emph{unital}, and indeed the algebra of \inlinecoq{rmap}s has all of these properties\footnote{In prior versions of VST, the separation algebra was also assumed to be \emph{cancellative}, which is not true of some useful forms of ghost state. As part of our extension, we relaxed this requirement; the algebra of heap resources is still assumed to be cancellative, but the combination of heap and ghost state need not be. We then re-proved all the separation logic rules of Verifiable C without assuming cancellativity.}. We can show that as long as every ghost algebra in the \inlinecoq{ghost} of our extended \inlinecoq{rmap}s meets these requirements, then the induced separation algebra on the \inlinecoq{rmap}s themselves also meets the requirements, making \inlinecoq{rmap}s with ghost state a valid model for VST's separation logic. This immediately gives us definitions of the basic separation logic operators, and guarantees that they satisfy the usual introduction and elimination rules.

\subsection{Ghost State in the Logic}

\begin{figure}[htb]
\centering
\begin{mathpar}
\inference[\textsf{own\_alloc}]{\mathsf{valid}\ a}{\mathsf{emp} \Rrightarrow \exists g.\ \mathsf{own}\ g\ a\ \mathit{pp}}

\inference[\textsf{own\_dealloc}]{}{\mathsf{own}\ g\ a\ \mathit{pp} \Rrightarrow \mathsf{emp}}

\inference[\textsf{own\_op}]{a_1 \cdot a_2 = a_3}{\mathsf{own}\ g\ a_3\ \mathit{pp} = \mathsf{own}\ g\ a_1\ \mathit{pp} * \mathsf{own}\ g\ a_2\ \mathit{pp}}

\inference[\textsf{own\_valid\_2}]{}{\mathsf{own}\ g\ a_1\ \mathit{pp} * \mathsf{own}\ g\ a_2\ \mathit{pp} \Rrightarrow \exists a3.\ a_1 \cdot a_2 = a_3 \land \mathsf{valid}\ a3}


\inference[\textsf{own\_update}]{\mathsf{fp\_update}\ a\ b}{\mathsf{own}\ g\ a\ \mathit{pp} \Rrightarrow \mathsf{own}\ g\ b\ \mathit{pp}}

\end{mathpar}
\caption{The separation logic rules for ghost state}
\label{ghost-rules}
\end{figure}

The next step is to define the separation logic assertions that make use of ghost state: most importantly, the $\mathsf{own}$ assertion, which asserts that a thread owns a piece of ghost state.
We do this using the same construction as Jung et al.~\cite{iris-ground-up}: we first define a larger $\mathsf{Own}$ predicate that describes the entire index-to-ghost-state map of an \inlinecoq{rmap}, and then $\mathsf{own}$ is defined as ownership of a singleton map at the indicated index. Interestingly, while Iris makes the type of the total ghost state map $M$ a parameter, we hide the type inside the \inlinecoq{rmap} using the dependent pair construction described above, which allows us to introduce new types of ghost state dynamically in our logic: practically speaking, this means that we do not need to declare that a particular type of ghost state is going to be used (in Iris, this is done with \inlinecoq{inG} assumptions that state that the global ghost state includes the desired algebras) before introducing it in a proof. Aside from this small improvement, and VST's different approach to \inlinecoq{rmap}s and step-indexing, our construction parallels that of Iris.

We can then prove the inference rules for $\mathsf{own}$ shown in Figure~\ref{ghost-rules}.
In the assertion $\mathsf{own}\ g\ a\ \mathit{pp}$, $g$ is an identifier (analogous to a location in the heap), $a$ is an element of a ghost algebra, and $\mathit{pp}$ is a separation logic predicate. In Iris, the element $a$ can include predicates in its own right; in VST, because of simplifications to avoid universe inconsistencies, we instead keep the predicate component (if any) in a separate argument, so that we can e.g. store separation logic assertions in ghost state to implement global invariants, without being able to express fully general higher-order ghost state. Otherwise, our $\mathsf{own}$ assertion functions in the same way as that of Iris. The key property of $\mathsf{own}$ is that separating conjunction on it corresponds to the join operator $\cdot$ of the underlying algebra (see rule \textsf{own\_op}).
Furthermore, ghost state can be modified at any time via a \emph{frame-preserving update}: an element $a$ can be replaced with an element $b$ as long as any third party's ghost state $c$ that is consistent with $a$ is also consistent with $b$, formally expressed as $$\mathsf{fp\_update}\ a\ b \triangleq \forall c, a \cdot c \Rightarrow b \cdot c$$ where we write $a \cdot b$ to mean $\exists d.\ a \cdot b = d$, i.e., $a$ and $b$ are compatible pieces of ghost state. This frame-preserving update is embedded into the logic via the \emph{view-shift} operator $\Rrightarrow$, as shown in rule \textsf{own\_update}. In the next section, we will extend VST's program logic to allow us to do view shifts at any point in a program, so that we can create, modify, and destroy ghost state as necessary in our verifications.

\subsection{Soundness of VST with Ghost State}

The last step in adding ghost state to VST is to redefine Hoare triples and re-prove their soundness. VST's Hoare triples are defined in terms of an inductive \emph{safety} property, as follows:
\begin{definition}[Safety]
A configuration $(c, h)$ is \emph{safe for} $n$ \emph{steps with postcondition} $Q$ if:

\begin{itemize}
\item $n$ is 0, or
\item $c$ has terminated and $Q(h)$ holds to approximation (step-index) $n$, or
\item $(c, h) \rightarrow (c', h')$ and $(c', h')$ is safe for $n - 1$ steps with $Q$.
\end{itemize}
\end{definition}
$\{P\}\ c\ \{Q\}$ then means that $\forall n\ h.\ P(h) \Rightarrow (c, h)$ is safe for $n$ steps with $Q$. We redefine this notion of safety to include ghost state that can undergo arbitrary frame-preserving updates between steps.
\begin{definition}[Safety with Ghost State]
A configuration $(c, h, g)$ is safe for $n$ steps with postcondition $Q$ if:

\begin{itemize}
\item $n$ is 0, or
\item $c$ has terminated and $Q(h, g)$ holds to approximation $n$, or
\item $(c, h) \rightarrow (c', h')$ and $\forall g_{\mathrm{frame}}.\ g \cdot g_{\mathrm{frame}} \Rightarrow \exists g'.\ (g' \cdot g_{\mathrm{frame}} \wedge (c', h', g')$ is safe for $n - 1$ steps with $Q)$.
\end{itemize}
\end{definition}
After each program step, we quantify over a frame ghost state $g_{\mathrm{frame}}$ consistent with our current ghost state, which represents ghost state held by other threads, the outside world, etc. We can then update our own ghost state to any $g'$ that is also consistent with the frame, and show that the rest of the program is safe with $g'$.

We then re-prove each of the separation logic rules of Verifiable C (for assignment, load, store, function calls, etc.) using this new definition of safety, as well as proving the usual enhanced rule of consequence:
$$\inference{P \Rrightarrow P' \qquad \{P'\}\ C\ \{Q'\} \qquad Q' \Rrightarrow Q}{\{P\}\ C\ \{Q\}}$$
In standard VST verifications, the rule of consequence is often applied via a tactic \inlinecoq{replace_SEP} that allows the user to replace a separating conjunct of the precondition with a new assertion entailed by that conjunct; accordingly, we define a new tactic \inlinecoq{viewshift_SEP} that replaces a conjunct with an assertion that follows from it via a view shift. We also provide a tactic \inlinecoq{ghost_alloc} that creates new ghost state with the requested initial value, making the use of ghost state in proof scripts as seamless as possible.

\subsubsection{Example: Parallel Increment}

\begin{figure}[htb]
\begin{subfigure}[t]{0.5\textwidth}
\begin{lstlisting}[style=customc]
lock_t ctr_lock;
lock_t thread_lock;
unsigned ctr;

void incr() {
  acquire(&ctr_lock);
  ctr = ctr + 1;
  release(&ctr_lock);
}

void thread_func() {
  lock_t *l = &thread_lock;
  incr();
  release(l);
}
\end{lstlisting}\end{subfigure}~
\begin{subfigure}[t]{0.5\textwidth}
\begin{lstlisting}[style=customc,numbers=none]
int main(void) {
  ctr = 0;
  makelock(&ctr_lock);
  release(&ctr_lock);
  makelock(&thread_lock);
  spawn(&thread_func);

  incr();

  acquire(&thread_lock);
  acquire(&ctr_lock);
  // ctr should be 2
  freelock(&thread_lock);
  freelock(&ctr_lock);
}
\end{lstlisting}\end{subfigure}
\caption{Parallel increment of a locked counter}
\label{incr}
\end{figure}

We illustrate the use of ghost state in VST with the classic parallel increment example (code shown in Figure~\ref{incr}). The \lstinline{ctr} variable is only modified when \lstinline{ctr_lock} is held, and is incremented by two threads (\lstinline{main} and a spawned thread running \lstinline{thread_func}) before being reclaimed. Without ghost state, we can prove that this program is safe (using a lock invariant $\exists n.\ \texttt{ctr} \mapsto n$), but not that \lstinline{ctr} is 2 at the end of the program (because the invariant does not allow threads to record changes to \lstinline{ctr}). We overcome this in the traditional way, by adding ghost variables shared between each thread and the lock invariant, where the lock invariant maintains the fact that the current value is the sum of the contributions: $R \triangleq \exists n.\ (n = n_1 + n_2) \land \texttt{ctr} \mapsto n * \mathsf{ghost\_var}_{.5}\ n_1\ g_1 * \mathsf{ghost\_var}_{.5}\ n_2\ g_2$, where $\mathsf{ghost\_var}$ is the $\mathsf{own}$ predicate instantiated with an algebra of share-annotated values. The heart of the proof is the specification of \lstinline{incr}: $$\{\texttt{ctr\_lock} \islock R * \mathsf{ghost\_var}_{.5}\ n\ g_i\}\ \texttt{incr}()\ \{\texttt{ctr\_lock} \islock R * \mathsf{ghost\_var}_{.5}\ (n + 1)\ g_i\}$$
Every time we increment \lstinline{ctr}, we also increment one of the ghost variables, so once \lstinline{main} rejoins with the spawned thread and collects its ghost variable, we can tell that \lstinline{ctr} is exactly 2. The new features of VST show up in only a few places in the proof: the $\mathsf{ghost\_var}$ predicate in the lock and \lstinline{incr} specifications, a \inlinecoq{viewshift_SEP} tactic in the proof of \lstinline{incr} that updates the ghost state (using the fact, derived from \textsf{own\_update}, that $\mathsf{ghost\_var}_1\ n\ g \Rrightarrow \mathsf{ghost\_var}_1\ (n + 1)\ g$), and two uses of the \inlinecoq{ghost_alloc} tactic in the proof of \lstinline{main} to initialize the ghost variables. Because the extensions described so far are foundational, we obtain VST's usual soundness guarantee for the program: when we compile it with CompCert and run it, the value of \lstinline{ctr} will actually be 2.

\subsection{Defining Invariants}
\label{invariants}
One of the most surprising and elegant results of Jung et al.~\cite{iris-ground-up} is that invariants, initially presented as one of the main primitives of the Iris logic, can in fact be constructed out of higher-order ghost state. Although VST's ghost state is not fully higher-order, we demonstrate in this section that it is sufficient to construct invariants, giving us access to one of the most powerful and versatile reasoning techniques in modern concurrent separation logic.

The earliest revisions of concurrent separation logic~\cite{gotsman,oracle-semantics} centered on the use of \emph{lock invariants}, where each lock is associated with the resources it protects. A lock invariant is expressed as an assertion $\ell \islock R$, where $\ell$ is the location of the lock and $R$ is an arbitrary separation logic assertion describing resources that can only be accessed by a thread that holds the lock. The rules governing the invariant are generally of the form:
\begin{mathpar}
\inference{}{\{\ell \islock R\}\ \texttt{acquire}(\ell)\ \{\ell \islock R * R\}}

\inference{}{\{\ell \islock R * R\}\ \texttt{release}(\ell)\ \{\ell \islock R\}}
\end{mathpar}
In combination with ghost state, these rules can be used to prove many useful correctness properties of lock-based programs. 
However, they also enforce a one-to-one association between synchronization object (the lock $\ell$) and protected resource (the invariant $R$). We can use ghost state to develop more complicated patterns, but if we have a program where threads can write to a pointer when they hold either of two locks (with some other protocol ensuring that both locks are not in use at the same time), or one that uses an entirely lock-free synchronization mechanism, we will struggle to specify it using lock invariants. Iris's solution is to generalize lock invariants away from locks entirely: an \emph{invariant} is simply a property that is true before and after every step of execution. We write $\knowInv{}{I}$ to assert that $I$ is an invariant for the current program. We can recover lock invariants with a property like $\knowInv{}{(\ell \mapsto 0 * R) \vee (\ell \mapsto 1)}$, which says that either the lock is not held by any thread (value 0) and the resources $R$ are owned by the invariant, or the lock is held (value 1) and, implicitly, $R$ is owned by some thread.

Invariants are constructed as a form of higher-order ghost state using the \emph{agreement} ghost algebra, in which elements are assertions of the separation logic and two elements join if they express the same predicate. In fact, the algebra in Iris is somewhat more complicated---it can be used to capture agreement on any type, not just assertions, and is carefully designed to respect step-indexing and avoid circular reference. We define a more constrained version in VST, for the sole purpose of defining invariants. In fact, agreement is already baked into our $\mathsf{own}$ assertion, since $\mathsf{own}\ g\ a\ p * \mathsf{own}\ g\ b\ q$ can only hold when $p = q$, so we can define $\mathsf{agree}\ g\ R$ as simply $\mathsf{own}\ g\ \mathsf{tt}\ R$. 

The core of the invariant construction is a ``world satisfaction'' assertion $\mathsf{wsat}$ that captures the contents and current state of every invariant in a program. Our construction is slightly modified from that of Iris, and consists of three parts: 
\begin{itemize}
\item A list of ghost names $g_i$ for all defined invariants, wrapped in the master-snapshot algebra~\cite{gps-rcu} so that threads can remember the fact that a particular name has been defined: $\mathsf{wsat}$ holds a master copy of the list, and any thread with access to it can create a $\mathsf{snapshot}$ recording the existence of one or more invariant names.
\item A set of ``enabled'' tokens and a set of ``disabled'' tokens, one for each defined invariant. For each invariant, exactly one of these two is always present in $\mathsf{wsat}$; an invariant is considered \emph{disabled} when $\mathsf{wsat}$ holds its \emph{enabled} token (and the thread modifying its contents holds its \emph{disabled} token), and vice versa.
\item For each invariant $i$, either $\mathsf{agree}\ g_i\ R_i$ if it is disabled, or $R_i * \mathsf{agree}\ g_i\ R_i$ if it is enabled.
\end{itemize}
We can then define $\knowInv{g}{R}$ as $\mathsf{snapshot}\ g * \mathsf{agree}\ g\ R$, i.e., the knowledge that $g$ is the name of some invariant and $R$ is the assertion in that invariant. In Iris, the first and third items are collapsed into a single piece of agreement ghost state that holds the full map from names to assertions, and an invariant assertion is a snapshot of that map; our more complicated construction is a workaround for the fact that our agreement algebra can only capture assertions, not maps of assertions. The rules for allocating invariants and the definition of ``fancy'' updates (ghost updates that open and close invariants) are derived as in Iris. 

So far, we have \emph{reimplemented} the ghost state and invariants of Iris in VST rather than importing their definitions. This is necessary, since we have made slight modifications to the foundational model, and also useful, since it means that the typical VST user can still build the toolchain (and even do interesting proofs about concurrent programs) without installing or understanding Iris. Users looking to verify sophisticated fine-grained concurrent programs, however, may well be familiar with Iris already and want to take advantage of its features. In the next section, we stop retracing Iris's steps and start integrating Iris directly into VST.

\section{Iris Proof Mode in VST}
Iris Proof Mode (IPM) provides separation-logic analogues of Coq tactics---\texttt{iDestruct}, \texttt{iAssert}, etc.---so that users can manipulate separation logic hypotheses and goals in more or less the same way as pure Coq propositions. The most recent iteration of IPM, called MoSeL~\cite{mosel}, decouples this system from the base logic of Iris by providing a typeclass of logics of bunched implications (BI): any logic that supports the basic rules of a BI logic, either as axioms or derived properties, can be targeted by the IPM/MoSeL tactics. In this section, we briefly summarize this interface, and then describe how we show that VST's logic is an instance of it, allowing us to use Iris tactics to prove separation logic assertions in VST.

The first stage of the interface, called \texttt{BiMixin}, consists mostly of the usual properties of logical connectives: $\wedge$, $\vee$, $\forall$, $\exists$, and the separating conjunction $*$ and implication $\wand$. There are also two more specialized operators: $\ulcorner P \urcorner$, which injects a pure Coq proposition $P$ into the separation logic, and the \emph{persistence modality} $\pers$. The persistence operator must satisfy the rule $\pers P \vdash P * \pers P$, i.e., a persistent assertion can be duplicated any number of times. A persistent assertion may simply not take up any resources in the state---like the empty heap assertion $\mathsf{emp}$ or a pure proposition---but it may also refer to a resource that is inherently duplicable, like ghost state asserting that some operation has occurred. Because assertions under the $\pers$ operator are arbitrarily duplicable and interact well with other operators, they are exempt from the usual linearity requirements of separation logic: if we learn $\pers P$ in the process of proving $Q * R$, then we can use it in proving both $Q$ and $R$. Because of this, MoSeL treats persistent assertions more like pure Coq propositions than separation logic assertions: they are kept in a separate list of hypotheses that do not need to be explicitly invoked or allocated to particular branches in the course of a proof.

The implementation of $\pers$ in the Iris logic is based on the \emph{core} operation of its resource algebras: each element $a$ of an algebra may have a core element $|a|$ such that $|a| \cdot a = a$, and $\pers P$ holds on $a$ when $P$ holds on $|a|$. Using this definition, many common features of CSL proofs---invariants, snapshots of ghost state, protocol assertions for weak memory---can be shown to be persistent, and so qualify for special treatment in the proof mode. VST also has a notion of core in its separation algebras, but the notion is slightly different. In particular, VST requires that if $a \preceq b$ then $|a| = |b|$  (where we write $a \preceq b$ to mean $\exists c.\ a \cdot c = b$), while in Iris $a \preceq b$ only implies that $|a| \preceq |b|$. We can see the implications of this difference, for instance, in an algebra of states in a finite state machine: if there is a transition $q_1 \rightarrow q_2$ in the state machine, then in Iris it may be the case that $|q_2| = q_2$, while in VST we must have $|q_2| = |q_1|$. If we interpret $|q|$ as ``the knowledge that we have at least reached state $|q|$'', the Iris core provides significantly more information than the VST core, allowing us to remember facts that were not true when we started the state machine but are guaranteed to be true from now on. Both notions of core are based on the work of Pottier~\cite{core}, but Iris intentionally weakens Pottier's axioms to gain more useful cores, as acknowledged in section 9.3 of Jung et al.~\cite{iris-ground-up}. The upshot is that VST's core can be used to define a $\pers$ modality that satisfies all the axioms of \texttt{BiMixin}, but invariants and many other ghost state constructs are not persistent in this instance. As a result, while we can prove that VST's logic is an instance of the BI typeclass, we do not get the full utility of the tactics when it comes to persistent hypotheses. We have not yet determined whether it is possible to weaken VST's core axiom without invalidating the rest of the separation logic.

The second stage of MoSeL's interface, called \texttt{BiLaterMixin}, introduces the ``later'' operator $\later$ and its properties. The $\later$ operator is a common feature of step-indexed systems, used to ensure that recursive constructions are well founded. Fortunately, Iris and VST are both built on the same concept of step-indexing (due to Appel and McAllester~\cite{step-index}), and VST's $\later$ operator satisfies all of the same properties. Once we have shown that VST's operators satisfy the properties of both stages, we gain full access to the MoSeL tactics in VST, with the above-mentioned caveat that assertions that are persistent in Iris cannot always be treated as persistent in VST.

It is worth noting that there are two layers of tactics used in Iris proofs of Hoare triples for programs: the MoSeL tactics \texttt{iDestruct}, \texttt{iAssert}, etc. for proving separation logic implications in the style of Coq, and tactics like \texttt{wp\_rec}, \texttt{wp\_if}, etc. that do symbolic execution on programs in the target language. VST has its own symbolic execution tactics (\texttt{forward}, \texttt{forward\_if}, etc.) which are specialized to C, have similar effects, and are arguably more powerful, so there is no reason to port these tactics to VST. We use Iris tactics only to prove separation logic entailments, such as those that arise when proving loop invariants and function preconditions. They are particularly useful for complicated implications involving multiple modalities and changes to ghost state, which are common when reasoning with invariants and especially with logical atomicity. In the next section, we examine atomicity proofs in VST in more detail.

\section{C11 Atomics and Logical Atomicity}
\subsection{Accessing Invariants with Atomic Operations}
\label{atomics}
Iris allows two main ways of interacting with established invariants. The first is to open and close an invariant in between steps of execution. 
The second and more powerful is to open an invariant for the duration of an \emph{atomic} step of execution, using a rule of the form\footnote{The actual rule in Iris uses fancy updates to allow opening any number of named invariants around $e$, while only opening each invariant at most once.}:
$$\inference{\mathsf{atomic}(e) \qquad \{\later I * P\}\ e\ \{\later I * Q\}}{\{\knowInv{}{I} * P\}\ e\ \{\knowInv{}{I} * Q\}}$$
In base Iris, $\mathsf{atomic}$ simply means that the expression computes in a single step of the underlying small-step semantics, and so ordinary memory loads and stores are considered to be atomic. This is distinct from the notion of atomicity in weak memory models, in which distinguished atomic operations perform synchronized memory access, while concurrent access through ordinary loads and stores leads to data races and possibly undefined behavior. For example, in base Iris, the program $p \leftarrow 3\ ||\ p \leftarrow 4$ could be successfully verified (using an invariant such as $p \mapsto 3 \vee p \mapsto 4$), while the corresponding program in C/Rust/etc. has undefined behavior and cannot guarantee that $p$ is either 3 or 4. In weak-memory Iris developments~\cite{iris-weak,rust-relaxed}, this is addressed by working with a more complex model of memory in which ordinary points-to assertions (referred to as ``non-atomic'', although they are still $\mathsf{atomic}$ in the sense of the rule above) are implicitly tagged with per-thread views of memory, and threads wishing to access them through invariants must explicitly reason about views in order to ensure sufficient synchronization.

VST's concurrent soundness proof is closely tied to CompCert's memory model~\cite{cpm}, and modifying it to support relaxed-memory reasoning will be a significant undertaking. In the meantime, we support \emph{sequentially consistent} (SC) atomic memory accesses (plus ordinary non-synchronized accesses), which allows a much simpler model. Invariants can be accessed using the rule above, except that $\mathsf{atomic}$ is only true of those operations that are atomic in the C sense, i.e., the built-in operations $\texttt{atomic\_load}$, $\texttt{atomic\_store}$, etc. As per the C11 specification, these operations can only be performed on pointers that have been declared to have an \texttt{atomic} type, and so we introduce a special points-to assertion $p \mapsto_a v$ for atomic pointers. We have not yet updated VST's soundness proof for this model---it will require adding the atomic operations to the Concurrent Permission Machine that serves as an interface between concurrent VST and the sequential semantics of CompCert---but we do not expect it to be significantly more difficult than the original soundness proof (in which the only synchronization operations are lock acquire and release). In the meantime, we can reason about C programs that use non-atomic and SC atomic memory accesses with a fairly high degree of confidence, and allow SC atomics to access Iris-style invariants. Instantiating the atomic rule with specific atomic operations gives us rules such as:
\begin{mathpar}
\inference{P * I \Rrightarrow \exists v.\ p \mapsto_a v * (x = v * p \mapsto_a v \Rrightarrow I * Q(v))}{\{\knowInv{}{I} * P\}\ x = \texttt{atomic\_load}(p)\ \{\knowInv{}{I} * Q(v)\}}

\inference{P * I \Rrightarrow p \mapsto_a \_ * (p \mapsto_a v \Rrightarrow I * Q)}{\{\knowInv{}{I} * P\}\ \texttt{atomic\_store}(p, v)\ \{\knowInv{}{I} * Q\}}
\end{mathpar}
If the points-to assertion $p \mapsto_a v$ is somewhere inside an invariant $I$, then we can access $p$ with an atomic operation, read or modify its value, and then return the assertion, re-establishing the invariant and retaining any leftover resources in the postcondition $Q$. The nested view shifts in the premises of these rules are exactly the sort of goal that is most easily proved with the IPM/MoSeL tactics described in the previous section.

\subsection{Logical Atomicity in VST}
The true power of Iris's invariants comes from the extension of atomicity to \emph{logical atomicity}. Logical atomicity originally appeared in the TaDA logic~\cite{tada} in the form of \emph{atomic triples} $\langle P \rangle\ c\ \langle Q \rangle$, which were then defined in Iris in terms of view shifts~\cite{iris}. If a function is proved to satisfy a logically atomic specification, then its effects are guaranteed to appear to take effect at a single point in time, and other threads will never observe an intermediate state of the function. This means that logically atomic functions can be treated as $\mathsf{atomic}$ in the sense above, and can access the contents of invariants. Logical atomicity gives us canonical specifications for data structure operations that can then be specialized with different ghost state and invariants depending on the task at hand: for instance, if a data structure is linearizable, we can combine its atomic specifications with an invariant that its current state is derivable from a sequential history of observed operations, and then do linearizability-based reasoning in a client proof.

The general form of an atomic triple is $\langle a.\ P_l\ |\ P_p(a) \rangle\ c\ \langle Q_l\ |\ Q_p(a)\rangle$, where $P_l$ and $Q_l$ are the \emph{local} or \emph{private} pre- and postconditions and $P_p$ and $Q_p$ are the \emph{public} pre- and postconditions. The public pre- and postconditions describe the state of an abstract object $a$ that can vary freely until the linearization point of $c$, at which point it must transition from a state satisfying $P_p$ to a state satisfying $Q_p$. (Note that despite its placement, the quantification of $a$ scopes over both the postcondition and the precondition.)
For example, an atomic enqueue operation for a queue might be specified as $\langle q.\ \mathsf{IsQueue}(p)\ |\ \mathsf{Queue}(p, q) \rangle\ \mathrm{enqueue}(p, v)\ \langle \mathsf{IsQueue}(p)\ |\ \mathsf{Queue}(p, v :: q) \rangle$, asserting that the caller must hold a reference to the queue at location $p$ ($\mathsf{IsQueue}(p)$), and enqueue atomically updates the contents of that queue from $q$ to $v :: q$ at some point during the execution of the function.

In Iris, atomic triples are derived forms, defined as:
$$\langle a.\ P_l\ |\ P_p(a) \rangle\ c\ \langle Q_l\ |\ Q_p(a)\rangle \triangleq \forall Q.\ \{P_l * \mathsf{atomic\_shift}(P_p, Q_p, Q)\}\ c\ \{Q_l * Q\}$$
where $\mathsf{atomic\_shift}$ is defined as roughly\footnote{The precise definition in Iris has gone through several revisions. The one presented here is a simplified version of the definition from the original Iris paper~\cite{iris}.}:
$$\mathsf{atomic\_shift}(\alpha, \beta, Q) \triangleq \exists P.\ \later P * \square (\later P \Rrightarrow \exists x.\ \alpha(x) * ((\alpha(x) \Rrightarrow \later P) \land (\beta(x) \Rrightarrow Q)))
$$
This is a rather complicated definition, but it captures the essence of an atomic update to the abstract state: we have a black-box resource $P$ that can be used to temporarily obtain the public precondition $\alpha$ on some state $x$, but we must always either make no visible changes and restore $\alpha(x)$, or (at the linearization point) satisfy the public postcondition $\beta$ and obtain $Q$ instead. When we prove that a function satisfies an atomic specification, $P$ and $Q$ are black boxes, and so the \emph{only} way to obtain $Q$ in the postcondition is by satisfying the public postcondition $Q_p$ at some linearization point during the function. When we call a function with an atomic specification, we can give it any arbitrary postcondition $Q$, as long as that postcondition can be shown to result from an atomic shift from $P_p$ to $Q_p$. 

Embedding this definition in VST's notation for function specifications is tricky in practice but simple in theory. A higher-order function specification in VST has the form
$$\begin{array}{l}
\texttt{TYPE}\ a\ \texttt{WITH}\ x_1, ..., x_n\\
\texttt{PRE}\ [ \mathit{ty}_1, ..., \mathit{ty_k} ]\\
\quad\quad\texttt{PROP}\ (P)\ \texttt{PARAMS}\ (R)\ \texttt{GLOBALS}\ (G)\ \texttt{SEP}\ (S)\\
\texttt{POST}\ [ \mathit{ty} ]\\
\quad\quad\texttt{PROP}\ (P')\ \texttt{RETURN}\ (R')\ \texttt{SEP}\ (S')
\end{array}$$
where $x_1$, ..., $x_n$ are parameters of the specification whose type is given by $a$, $\mathit{ty}_1, ..., \mathit{ty}_k$ and $\mathit{ty}$ are the C types of the function's arguments and return value, and the pre- and postcondition are broken up into pure Coq assertions (\texttt{PROP}), assertions about program variables (\texttt{PARAMS}/\texttt{GLOBALS}/\texttt{RETURN}), and separation logic assertions (\texttt{SEP}). For logically atomic specifications, we define a notation
$$\begin{array}{l}
\texttt{ATOMIC TYPE}\ a\ \texttt{OBJ}\ o\ \texttt{WITH}\ x_1, ..., x_n\\
\texttt{PRE}\ [ \mathit{ty}_1, ..., \mathit{ty_k} ]\\
\quad\quad\texttt{PROP}\ (P)\ \texttt{PARAMS}\ (R)\ \texttt{GLOBALS}\ (G)\ \texttt{SEP}\ (S)\ |\ (S_p)\\
\texttt{POST}\ [ \mathit{ty} ]\\
\quad\quad\texttt{PROP}\ (P')\ \texttt{RETURN}\ (R')\ \texttt{SEP}\ (S')\ |\ (S_p')
\end{array}$$
where the abstract object $o$ corresponds to the $a$ in $\langle a.\ P_l\ |\ P_p(a) \rangle\ c\ \langle Q_l\ |\ Q_p(a)\rangle$, $S_p$ and $S_p'$ correspond to $P_p$ and $Q_p$ respectively, and the rest of the conditions correspond to $P_l$ and $Q_l$. 
This notation is desugared into
$$\begin{array}{l}
\texttt{TYPE}\ a * \texttt{mpred}\ \texttt{WITH}\ x_1, ..., x_n, Q\\
\texttt{PRE}\ [ \mathit{ty}_1, ..., \mathit{ty_k} ]\\
\quad\quad\texttt{PROP}\ (P)\ \texttt{PARAMS}\ (R)\ \texttt{GLOBALS}\ (G)\ \texttt{SEP}\ (\mathsf{atomic\_shift}(S_p, S_p', Q) * S)\\
\texttt{POST}\ [ \mathit{ty} ]\\
\quad\quad\texttt{PROP}\ (P')\ \texttt{RETURN}\ (R')\ \texttt{SEP}\ (Q * S')
\end{array}$$
In other words, we add the desired postcondition $Q$ as another parameter to the specification, convert the public pre- and postcondition into an $\mathsf{atomic\_shift}$ added to the ordinary precondition, and add $Q$ to the postcondition, mirroring the Iris definition of atomic triples. (This notation was quite hard to implement, especially the extension of the type of the \texttt{WITH} clause, and benefited from the advice of Jason Gross.)

This gives us a way to write atomic function specifications in the usual VST style, after which we can prove that they are satisfied by implementations, and use them in client programs, with the existing tactics of VST. The key differences are that 1) when proving that a function meets an atomic specification, we can only access the resources in $S_p$ via the atomic shift, and must then either restore it as is or, at a linearization point, obtain $Q$ by satisfying the public postcondition $S_p'$; and 2) when calling a function with an atomic specification, we must provide the desired postcondition $Q$ as an extra parameter, and prove that the $\mathsf{atomic\_shift}$ from $S_p$ to $S_p'$ causes $Q$ to hold. By construction, we can access the contents of invariants when proving the atomic shift, thus fulfilling the promise of atomic triples: when calling a function with an atomic specification, we can access invariants around it as if it were a single atomic instruction. In the next section, we will illustrate this with a concrete example.

\subsection{Example: The World's Simplest Lock-Free Hash Table}
With ghost state, invariants, atomic memory operations, and logically atomic specifications at our disposal, we can now prove strong specifications for C implementations of lock-free data structures in VST. In this section, we consider a C adaptation of Preshing's World's Simplest Lock-Free Hash Table~\cite{hashtable}. The hashtable is implemented as an array of key-value pairs, where both keys and values are atomically accessed integers (type \lstinline{atom_int}). The code for the \lstinline{set_item} function is shown in Figure~\ref{hashtable-set}. We hash the target key \lstinline{key} to get an index \lstinline{idx} into the array, and atomically read the key at \lstinline{idx} (line 5) to see if it it matches our \lstinline{key}. If it does, we atomically store the target value \lstinline{value} at \lstinline{idx} (line 20). If there is some other key at \lstinline{idx}, we ``chain'' by simply incrementing \lstinline{idx} by 1 and trying again (line 10). If we find an empty slot (indicated by key 0), we attempt to CAS it to \lstinline{key} (line 11). If the CAS fails (i.e., the key at \lstinline{idx} has been set since our initial read), then we check again whether the current key is equal to \lstinline{key}, and either set the value or move to the next index as appropriate. The \lstinline{get_item} function (Figure~\ref{hashtable-get}) simply moves through the keys until it finds either the target key (in which case it returns the associated value), or an empty entry (in which case the key is not in the table). We also include an \lstinline{add_item} function that adds a key only if it is not found in the table.

\begin{figure}[htb]
\begin{lstlisting}[style=customc]
void set_item(int key, int value){
  int ref = 0;
  for(int idx = integer_hash(key);; idx++){
    idx &= ARRAY_SIZE - 1;
    atom_int *i = m_entries[idx].key;
    int probed_key = atom_load(i);
    if(probed_key != key){
      //The entry was free, or contained another key.
      if (probed_key != 0)
        continue;
      int result = atom_CAS(i, &ref, key);
      if(!result){
        //CAS failed, so a key has been added. Is it our key?
        probed_key = atom_load(i);
        if(probed_key != key)
          continue; //This slot has been taken for another key.
      }
    }
    i = m_entries[idx].value;
    atom_store(i, value);
    return;
  }
}
\end{lstlisting}
\caption{The \texttt{set} function for the lock-free hashtable}
\label{hashtable-set}
\end{figure}

\begin{figure}[htb]
\begin{lstlisting}[style=customc]
int get_item(int key){
  for(int idx = integer_hash(key);; idx++){
    idx &= ARRAY_SIZE - 1;
    atom_int *i = m_entries[idx].key;
    int probed_key = atom_load(i);
    if(probed_key == key){
      i = m_entries[idx].value;
      return atom_load(i);
    }
    if (probed_key == 0)
      return 0;
  }
}
\end{lstlisting}
\caption{The \texttt{get} function for the lock-free hashtable}
\label{hashtable-get}
\end{figure}

To specify the hashtable functions, we begin by describing the relationship between the abstract state of the hashtable as a map from keys to values, and its concrete state as a collection of atomically accessed integers. An individual hashtable entry is a pair of atomic integers $\pk, \pv$ with values $k, v$ and a ghost state $g$, defined as follows:
$$\mathsf{entry}\ (\pk, \pv)\ (k, v)\ g \triangleq (k = 0 \rightarrow v = 0) \land \pk \mapsto_a k * \pv \mapsto_a v * \mathsf{ghost\_master}\ k\ g$$
Each entry maintains the invariant that if its key is empty, its value is also empty. We also include a piece of ghost state from the master-snapshot pattern~\cite{gps-rcu} with value $k$, with a simple order in which 0 is less than any non-zero number. This allows a client to take a snapshot that records the key it has seen at an entry, with the knowledge that any non-zero key will remain at that entry indefinitely (but a 0 value may be replaced with a non-zero key). A hashtable of size $\mathit{size}$ is simply a collection of these entries:
$$\mathsf{hashtable}\ H\ \vec{p}\ \vec{g} \triangleq \exists \vec{e}, (\vec{e} \textrm{ implements } H) \land \Sep_{i \in \mathit{size}} \mathsf{entry}\ p_i\ e_i\ g_i$$
where $\vec{e}$ is the list of key-value pairs that appear in the table, and the abstract state of the hashtable is a Coq-level map from keys to values $H$ that is consistent with $\vec{e}$.

We can use a sequence of snapshots of the entries to precisely determine where a key should appear in the hashtable. Specifically, we prove the following lemma:
\begin{lemma}
\label{entries-lookup}
For any key $k$, if we have snapshots of the keys from entries with indices in the range $[\mathsf{hash}(k), \mathsf{hash}(k) + i)$ (modulo size) such that none of the snapshots are either 0 or $k$, and the key at index $\mathsf{hash}(k) + i$ is either 0 or $k$, then $\mathsf{hash}(k) + i$ is the unique index in the hashtable where key $k$ can be found or inserted.
\end{lemma}
The proof involves a good deal of modular arithmetic, but the intuition is simple: because our chaining function is simply ``move to the next index'', the key $k$ will always appear at the end of a (possibly empty) sequence of non-empty, non-$k$ keys starting from the index $\mathsf{hash}(k)$.

Now we can state the specifications of the hashtable functions. On paper, these specifications are a fairly intuitive collection of logically atomic triples:
\[\begin{array}{c}
\langle H.\ h \mapsto \vec{p}\ |\ k \neq 0 \land \mathsf{hashtable}\ H\ \vec{p}\ \vec{g}\rangle\\ \texttt{set\_item}(h, k, v)\ \\\langle h \mapsto \vec{p}\ |\ \mathsf{hashtable}\ H[k \mapsto v]\ \vec{p}\ \vec{g}\rangle
\\
\\
\langle H.\ h \mapsto \vec{p}\ |\ k \neq 0 \land \mathsf{hashtable}\ H\ \vec{p}\ \vec{g}\rangle\\ \texttt{get\_item}(h, k)\\ \langle v.\ h \mapsto \vec{p}\ |\ H(k) = v \land \mathsf{hashtable}\ H\ \vec{p}\ \vec{g}\rangle
\\
\\
\langle H.\ h \mapsto \vec{p}\ |\ k \neq 0 \land \mathsf{hashtable}\ H\ \vec{p}\ \vec{g}\rangle\\ \texttt{add\_item}(h, k, v)\\ \langle b.\ h \mapsto \vec{p}\ |\ (k \not\in H \land b = \texttt{true} \land \mathsf{hashtable}\ H[k \mapsto v]\ \vec{p}\ \vec{g}) \vee (k \in H \land b = \texttt{false} \land \mathsf{hashtable}\ H\ \vec{p}\ \vec{g})\rangle
\end{array}\]
The caller requires a reference to the location of the hashtable entries ($h \mapsto \vec{p}$), and each function performs its operation atomically on the abstract state $H$ of the hashtable. In practice, VST requires more annotations and side conditions, so that the specification of the \lstinline{get_item} function looks like:
\begin{lstlisting}[style=customcoq]
Program Definition get_item_spec := DECLARE _get_item
  ATOMIC TYPE (ConstType (Z * globals * share * list (val * val) * gname * list gname))
  OBJ H WITH k, gv, sh, entries, g, lg
  PRE [ tint ]
    PROP (readable_share sh; repable_signed k; k <> 0;
          Forall (fun '(pk, pv) => isptr pk /\isptr pv) entries;
          Zlength lg = size)
    PARAMS (vint k) GLOBALS (gv)
    SEP (data_at sh (tarray tentry size) entries (gv _m_entries)) | (hashtable H g lg entries)
  POST [ tint ]
   EX v : Z,
    PROP ()
    LOCAL (temp ret_temp (vint v))
    SEP (data_at sh (tarray tentry size) entries (gv _m_entries)) |
        (!!(if eq_dec v 0 then H k = None else H k = Some v) && hashtable H g lg entries).
\end{lstlisting}
The \lstinline[style=customcoq]{data_at} predicate represents the points-to assertion (the global variable \lstinline{m_entries} holds the hashtable entries, represented by the list of pointer pairs \lstinline{entries}). The caller need only hold a readable share of this assertion, so it can be divided among multiple threads. The key \lstinline[style=customcoq]{k} must be non-zero, but it must also be representable as a machine integer (i.e., it must be between the minimum and the maximum integer), a detail we would ignore in a less concrete language model.

The proof that \lstinline{get_item} satisfies this spec begins as a conventional VST proof, with the atomic shift as an extra SEP clause in the precondition, and the extra assertion $Q$ in the postcondition (representing the postcondition requested by the caller, and treated as a black box in the proof). The invariant for the main loop is that we have not passed the location of the target key or an empty slot, which we express through a list of snapshots of the ghost state for each key seen (initially empty), as described in Lemma~\ref{entries-lookup}. The new features of VST first come into play at the first atomic load at line 5. When we call \texttt{atomic\_load}, we also supply the desired postcondition for the load, which as seen in section~\ref{atomics} is a predicate on the returned value $v$. In this case, we have two possible results. If we read a 0, then we have reached the end of the hashtable without finding the key $k$, and this load is our linearization point: $k$ is not currently in $H$, and 0 is the correct return value. If we read a nonzero value, line 5 is not the linearization point, and we instead record a snapshot of the observed key (copied from $\mathsf{ghost\_master}$ in the $\mathsf{entry}$) and continue executing. So the postcondition we request is $Q_1 = (\lambda v.\ \textrm{if } v = 0 \textrm{ then } Q(v) \textrm{ else } P * \mathsf{ghost\_snap}\ v\ g_i)$, where $Q$ is the black-box assertion received from fulfilling the atomic shift at the linearization point. We pass this postcondition as an argument to the \texttt{forward\_call} tactic, and receive the premise of the \texttt{atomic\_load} rule as a proof obligation: $P * I \Rrightarrow \exists v.\ p \mapsto_a v * (x = v * p \mapsto_a v \Rrightarrow I * Q_1(v))$, where $I = \mathsf{hashtable}\ H\ \vec{p}\ \vec{g}$.

At this point, our goal is an implication between two separation logic formulae involving ghost state and atomic shifts---exactly where Iris Proof Mode is most useful. We switch into IPM simply by applying the \texttt{iIntros} tactic used to start Iris proofs, and now we have the features of Iris at our disposal: named premises, the \texttt{iMod} tactic for handling ghost updates, and so on. The proof of this implication proceeds entirely by Iris and base Coq tactics, with no VST-specific reasoning. When the key is 0, we use our snapshots and lemma~\ref{entries-lookup} to show that the target key $k$ must not be in the hashtable, and so we can discharge our atomic shift and obtain the black-box postcondition $Q$ with value 0 (key not found). Otherwise, we take a snapshot of the current entry's key and add it to our list of snapshots. Note that in the ``key not found'' case, we do not yet return from the function, but we have the postcondition $Q$ in our current state, representing the fact that we have reached the linearization point and will do no further atomic operations.

Once we finish the proof obligation of the \texttt{atomic\_load}, we are back in a VST proof state, and can resume using VST tactics to step through the C code. We have three cases left to consider. First, when \texttt{probed\_key == key}, we do another \texttt{atomic\_load} (line 8) to read the value associated with that key. This is always a linearization point, and the value we read will be the return value for the function, so the desired postcondition is simply $Q$. Again we shift into IPM to manage the ghost state and invariant operations, use the collected snapshots to show that the current index must be the unique location of the key $k$, and discharge the atomic shift to obtain $Q$. Second, when \texttt{probed\_key == 0}, we already have the postcondition $Q(0)$ from our earlier \texttt{atomic\_load}, and we can simply return 0 (line 11) and satisfy the postcondition of \lstinline{get_item}. Finally, if neither is the case, we continue to the next index and repeat the process. 

The complete verification of the hashtable consists of ~2kLoC in Coq, and is included in VST at \url{https://github.com/PrincetonUniversity/VST/blob/master/atomics/verif_hashtable_atomic.v}. In addition to the proofs of the hashtable operations, it includes a simple client that derives linearizability from the atomic specifications by associating the hashtable with a ghost history of operations, and demonstrates that if three threads each try to add the same three keys to the hashtable, then exactly three keys are added.
In summary, we use the extensions to VST in the following ways in the verification process:
\begin{enumerate}
\item In the specifications, we use ghost state to define the abstract state of the hashtable, and logical atomicity (the \texttt{ATOMIC} keyword) to describe the desired behavior of the data structure operations.
\item In the proofs of the functions with atomic specifications, we automatically receive an atomic shift describing the global state and the requirements on the linearization point, and a black-box postcondition $Q$ that we must obtain from the atomic shift in order to satisfy the function's postcondition.
\item When we call atomic memory operations like \texttt{atomic\_load}, we automatically make use of the new specifications for these operations. The preconditions of these operations include proof obligations that we use Iris Proof Mode to discharge: we open invariants to access the current global state of the data structure, update ghost state as needed, and then either restore the invariants or (at linearization points) satisfy the public postcondition. These proofs appear as IPM subsections within the VST proof scripts for the data structure operations, and we shift seamlessly back to VST tactics once the relevant subgoals are discharged.
\item Similarly, in proofs of clients of the data structure we use ordinary VST tactics to reason about the C code, and drop into IPM to prove the preconditions of functions with logically atomic specifications (like \lstinline{get_item}).
\end{enumerate}

\section{Conclusion}
We have now seen how several features of Iris can be variously reconstructed, incorporated, or axiomatized in VST. The result is a verification system for concurrent C programs that benefits from the advances of modern separation logic: custom ghost state, global invariants, Coq-style tactics for SL reasoning, and logically atomic specifications. More work remains to foundationally verify some of these features (specifically the proof rules for atomic operations and their interaction with global invariants), but for the time being we have an effective system for program verification whose unverified axioms have been shown to be sound in other (non-C-specific) logics. The specifications and proofs of this system appear as fairly simple extensions to the usual VST style, and undergraduate and masters students have been able to learn the system well enough to verify simple fine-grained concurrent data structures~\cite{roshan-thesis}.

Aside from closing the soundness gap, there are several areas for future extensions. Iris proofs make heavy use of the persistence modality, which allows ``information''-style resources like global invariants to be freely duplicated with minimal effort, but differences in the underlying separation algebra mean that most of these resources are not persistent in VST. We intend to investigate whether VST's ``core'' axiom can be weakened to match the one in Iris, which would give us the full power of persistence in VST. We are also very interested in doing weak-memory reasoning in VST, and Iris has already been used as a base for several powerful weak-memory logics~\cite{iris-weak,rust-relaxed}. These would involve much more careful analysis (if we want to add their rules as axioms) or larger changes to VST's foundations and soundness proofs (if we want to foundationally verify the rules), but the results would be well worth it---a system for directly proving correctness of real C implementations of weak-memory algorithms and data structures. 

\paragraph{Acknowledgements} With thanks to Robbert Krebbers and Joseph Tassarotti, for convincing me that Iris tactics were necessary for atomicity proofs; Shengyi Wang, for merging new features into the main branch of VST; and Jason Gross, for invaluable advice on notations for atomic specifications.

\bibliography{ref}

\end{document}